\documentclass[12pt]{article}
\usepackage{graphicx}
\usepackage[top=3cm, bottom=2cm, left=3cm, right=3cm]{geometry}
\title{Cosmic Rays in the Inner Galaxy and the Diffusion Properties of the Interstellar
 Medium}
\author{A. D. Erlykin $^{1,2}$ and A. W. Wolfendale $^{2}$\\
$(1)$ P N Lebedev Physical Institute, Moscow, Russia\\
$(2)$ Department of Physics, Durham University, Durham, UK}
\begin{document}
\maketitle

\begin{abstract}
Recent measurements of cosmic gamma ray intensities up to TeV energies have been used 
to estimate the spectral shape of the parent cosmic ray particles present in the 
interstellar medium. The case is made for the particle spectrum in the Inner Galaxy 
being flatter than locally and in the Outer Galaxy. Of various possible explanations we
 make the case for the propagation of the particles being different in the more 
turbulent interstellar medium of the Inner Galaxy. The characteristic parameter 
$\alpha$ for the so called 'anomalous diffusion' is expected to be less in the Inner 
Galaxy than that locally and the corresponding power law spectral exponent of the 
cosmic ray particles $\gamma$ will differ from that locally. Arguments are presented 
favouring a value of $\alpha$ lower than the local one by $\Delta \alpha$ = 0.2; the 
consequence is that $\Delta \gamma \sim$ 0.1 for the parent particles.
 \end{abstract}
\section{Introduction}
There is a wealth of information about those cosmic rays (CR) that are detected
 at Earth and satisfactory models involving acceleration by supernova remnants
(SNR) have been put forward, at least for energies to some few PeV. The CR in 
question (particles, and mainly protons) are thought to be generated by SNR within 
a few kpc of the Earth but  for more distant sources from which few, if any, particles 
arrive at Earth, other techniques must be used. 

Insofar as Neutrino Astronomy is only about to start, recourse is made to Gamma Ray 
Astronomy, the gamma rays being generated by CR particles interacting with gas and 
photons. Some information about distant SNR sources has come from the detection of 
gamma ray sources \cite{Funk}, although here there are problems in distinguishing 
between electron-Inverse Compton interactions and CR nuclei-gas interactions as the 
source of the gamma rays.

The purpose of the analysis described in this paper is to study whether there is a 
difference between CR energy spectra in the different parts of the Galaxy. It is known 
that there is
the spatial gradient of CR in the Galactic Disk which means that there are more CR in 
the Inner Galaxy than in the Outer Galaxy. However, this gradient is found to be small 
 and established only for low energy CR in the GeV region \cite{Dodd,Stro,Issa,Bloe}. 
Its existence at TeV energies is not known and we try to find it studying the CR energy
 spectrum in the Inner Galaxy in comparison with that locally and in the Outer  Galaxy.

The form of the paper is as follows. Experimental data are taken and corrected by 
removal of the extragalactic component and the contribution of discrete gamma ray 
sources. The remainder, which is the bulk of the measured intensity, is then compared 
with our model predictions and conclusions drawn about the mode of CR propagation.
\section{The data}  
There is a long history of cosmic gamma ray data from a series of satellites and ground
 based telescopes of
 ever improving collecting power and resolution. The satellite data used here for the
study of gamma rays in the Inner Galaxy are from Fermi LAT \cite{Acke}. In this paper 
 the Fermi LAT collaboration concluded that there is an excess of high energy gamma 
rays detected in the Inner Galaxy over the expectation assuming CR energy spectra there
 are the same as locally (~GALPROP model~). It means that the CR energy spectra in the 
Inner Galaxy are flatter than that locally. This conclusion was based on the 
analysis of the observed total energy spectrum including the contribution of discrete 
sources as well as the isotropic background. The analysed energy range was limited by 
the maximum energy of 0.1 TeV. 

Here, we examine not the total, but the so called diffuse component of gamma rays and 
choose regions of the Galaxy (~$30^\circ < \ell < 65^\circ, |b| < 2^\circ$~), where 
CR interactions 
with gas in the interstellar medium (ISM) predominate. The interactions are primarily 
between CR protons and the ISM, in which $\pi^\circ$-mesons are produced, the 
$\pi^\circ$ - decaying into two gamma rays each. In what follows these gamma 
rays are referred to as $\pi^\circ$-gamma rays. A non-negligible flux of gamma 
rays from CR electron interactions with photons and ISM gas (~Inverse Compton, 'IC' and
 bremsstrahling, 'BR'~) is also generated and is relevant. We subtract the 
contribution of the discrete sources where gamma rays are generated by the interactions
of accelerated CR (~protons, nuclei and electrons~) with the gas, photons and magnetic 
fields inside these sources ( SNR, pulsars etc ) using the published Fermi LAT 
estimates. We also subtracted the contribution of the isotropic gamma ray 
background, because it is mainly extragalactic and we are interested and restrict 
ourself only to the spatial distribution of the Galactic CR.   

In order to check the Fermi LAT indication of the gamma ray excess we introduced two 
new subjects: a) for the comparison of the observation with the expectation we used our
 own model of the CR acceleration and propagation (~anomalous diffusion model~) 
\cite{EW1,EW2} and b) to extend the analysed energy range we included experimental data
 of the MILAGRO collaboration which gave an estimate of the diffuse gamma ray flux at 
15 TeV \cite{Abd1,Abd2}.

MILAGRO is a ground based water-Cherenkov detector. The collaboration measured the 
gamma ray flux in the region with Galactic coordinates $30^\circ < \ell < 65^\circ,
|b| < 2^\circ$. Recognized 'discrete' gamma ray sources have been removed, these 
amounted to about 25\% of the flux. There are also data for the region with 
$65^\circ < \ell < 85^\circ, |b| < 2^\circ$ but we did not use them since this 
region is at the very far periphery of the Inner Galaxy and includes the intensive 
gamma ray source in the Cygnus region which was evidently not completely removed from 
the data. Unfortunately, the MILAGRO collaboration gave only the upper limits for the 
gamma ray flux from the Outer Galaxy which was rather high and not useful for our 
analysis.  
 
 The adopted Fermi LAT data are for the longitude and latitude ranges 
$|\ell| \leq 80^\circ$ and $|b| \leq 8^\circ$ respectively. Besides diffuse gamma rays 
they included contributions from discrete sources and isotropic background. We decided 
to use only diffuse gamma rays due to three reasons: a) as has been already 
mentioned, diffuse gamma rays are directly connected with Galactic CR via their 
interactions with the ISM; b) in the analysis of the data we used our calculations 
\cite{EW3} which were made for the diffuse component of the total gamma ray flux; 
c) the MILAGRO flux 
used for the comparison with Fermi LAT included only diffuse gamma rays. We removed 
from the published gamma ray flux contributions from detected gamma ray sources and 
isotropic background using their calculations with the GALPROP model 
$^SS^Z4^R20^T150^C5$ presented in the publication. The corrections were not big and 
reduced the total intensities by $\Delta(logI)=0.067$ at logE,GeV=-1 and by 
$\Delta(logI)=0.117$ at logE,GeV=2.   

In order to make a 
direct comparison between MILAGRO and Fermi LAT data the latter measurements were 
converted to the region of $30^\circ < \ell < 65^\circ, |b| < 2^\circ$ using our
calculations \cite{EW3}. The mean intensity in the latter region was found to be higher
compared with the former one. According to these calculations the conversion factor $R$
 was nearly independent of the energy, increasing smoothly from $logR = 0.434\pm0.002$ 
at logE,GeV = -1 to $0.494\pm0.004$ at logE,GeV = 2. Its uncertainty was calculated 
from the sampling errors of the mean intensity in the regions of $|\ell| < 
80^\circ, |b| < 8^\circ$ and $30^\circ < \ell < 65^\circ, |b| < 2^\circ$ averaged over 
$161 \times 9$ and $36 \times 3$ points separated by $\Delta \ell = 1^\circ$ and 
$\Delta b = 2^\circ$ respectively. Calculations included 50 samples of simulated 
space-time configurations of 50000 SNR. The shape of the experimental energy 
spectrum after the conversion has not been distorted significantly.
    
 The Fermi LAT collaboration also presented the gamma ray spectrum in
 the Outer Galaxy for $80^\circ \leq \ell \leq 280^\circ$, $|b| \leq 8^\circ$, which we
 also use here for comparison with our model ( see \S5 ). Since for the comparison of
the spectral shapes in the Inner and Outer Galaxy the absolute intensities are not 
important, we made a direct comparison of the data for the Inner (~$|\ell| < 80^\circ, 
|b| < 8^\circ$~) and Outer (~$80^\circ < \ell < 280^\circ, |b| < 8^\circ$~) Galaxy 
without any corrections.
     
The MILAGRO collaboration published the integral (~rather than the differential~) 
gamma ray intensity and since we are 
going to use both MILAGRO and Fermi LAT data to derive the spectral shape above the 
Fermi LAT maximum energy, we integrated the Fermi LAT intensities, too.  

\section{Analysis of the energy spectra of gamma rays}    
\subsection{Derivation of the spectra}
Figure 1 shows the integral spectrum for the region with Galactic coordinates  
$30^\circ < \ell < 65^\circ, |b| < 2^\circ$. The predictions come from our Monte
Carlo model of SNR acceleration of CR \cite{EW1,EW2} with the target gas, $HI$
and $H_2$ from the summary \cite{EW3}. 
\begin{figure}[ht]
\begin{center}
\includegraphics[height=15cm,width=10cm,angle=-90]{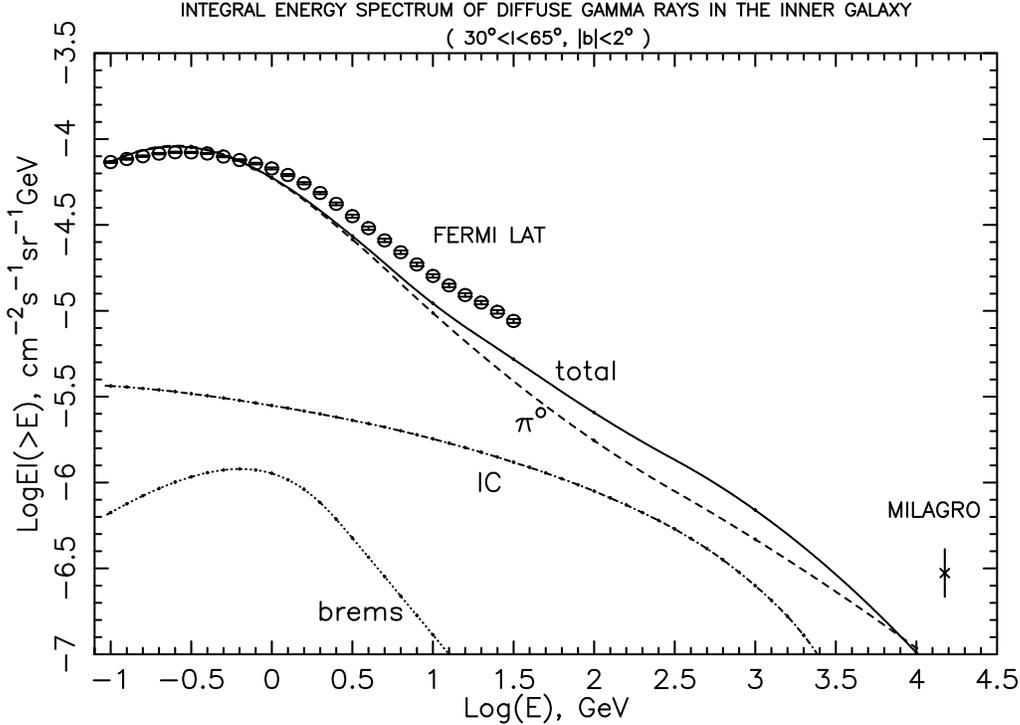}
\end{center}
\caption{\footnotesize Integral energy spectra of diffuse gamma rays in the Inner 
Galaxy. Experiment: open circles at $logE,GeV < 1.5$ - Fermi LAT measurements at
$|\ell| < 80^\circ, |b| < 8^\circ$, converted to the region covered by MILAGRO by
increasing the Fermi LAT intensities by the factor $\Delta log(EI(>E))$ which smoothly 
varies from 0.434 to 0.494 in the $logE,GeV$ interval from -1 to 2 respectively based 
on
calculations \cite{EW3}; the cross at $logE,GeV = 4.2$ is the MILAGRO measurement for 
the same region. Model calculations: dashed line -
$\pi^\circ$ - gamma rays ($\pi^\circ$), calculated for anomalous diffusion with 
$\alpha = 1$, dash-dotted line - gamma rays from Inverse Compton scattering of 
electrons on 
the interstellar radiation field (ISRF) photons (IC), dotted line - gamma rays from the
 bremsstrahlung of electrons on the ISM gas, full line - total diffuse gamma rays. The 
calculated total diffuse $\pi^\circ$ gamma ray intensity is 
increased to normalize it to the experimental one at $logE,GeV = -1$.}  
\label{fig:fig1}
\end{figure}

As for $\pi^\circ$-gamma rays, the proton injection spectrum with  differential energy 
spectrum exponent
of $\gamma = 2.15$ comes from our SNR acceleration model. Insofar as we 
normalize the prediction to observation at the lowest energy (~$logE,GeV = -1$~), no 
'metallicity correction' has been applied to the $H_2$ column densities. Anomalous
diffusion ('superdiffusion') has been adopted with the parameter $\alpha = 1.0$, which 
gives the best description of the CR spectrum observed locally \cite{EW4}. The case for
 anomalous diffusion is considered here and strengthened later, in \S4. 

Mathematical apparatus for anomalous diffusion of CR was developed by two groups in 
Russia
\cite{Lag1,Lag2,Ucha,Lag3}. It was applied for the analysis of various characteristics 
of CR (~energy spectra, mass composition, anisotropy, electrons and positrons etc.~) 
\cite{Lag4,Lag5,Lag6,Lag7,Lag8,Buga,Yush}. We prefer anomalous diffusion since it 
matches better the expected mode of propagation in the highly non-uniform ISM. It also 
helps to 
understand the formation of the huge Galactic Halo and the small radial gradient of CR 
in the Galaxy \cite{EW4,EW5} following from the analysis of Fermi LAT data \cite{Ferm} 
and previous measurements.
 The parameter $\alpha$ in the description of anomalous diffusion is  
determined by the structure of the turbulent ISM. It determines the temporal dependence
of the CR propagation and the shape of the diffusion front. The value $\alpha = 2$ is 
for the normal 'Gaussian' diffusion in the relatively uniform ISM with 
small turbulence, such as that in the far Outer Galaxy or in the Galactic Halo. The 
smaller $\alpha = 1$ value corresponds to an asymptotically faster diffusion, which 
fits better the conditions in the local environment (~and in the near Outer Galaxy~). 
The distinction between near and far Outer Galaxy arises because many of the 'Outer 
Galaxy' CR come from within a few kpc of the Sun and here the turbulence is still quite
 high ($\alpha = 1$) compared with our presumed $\alpha = 2$ for the far Outer Galaxy.

The connection of the turbulence spectra with the diffusion characteristics let us make
 some predictions about the steepening of the CR energy spectra with increasing  
Galactocentric radius, the rise of the radial gradient of CR at higher energies etc. 
Some of these predictions were already checked and confirmed by our analysis of EGRET 
data \cite{EW5}, others can be checked by further analysis of Fermi LAT data, and will 
be done so in what follows.    

Turning to the IC contribution, the electron spectrum in the Inner Galaxy was 
taken equal to the local one but truncated at 10 TeV, following \cite{EW2}. 
Interestingly, it is likely that because of the greater electron density losses in the 
Inner Galaxy (~due to the higher energy densities of the magnetic field and starlight 
there~) the electron cut-off energy will be significantly less than 10 TeV and the 
slight convexity of the 'total' in Figure 1 centred on $logE,GeV \simeq 2.5$ will be 
largely removed. In any 
event, this narrow range of latitudes leads to IC gamma rays comprising only a minority
 component. The photon intensities in the interstellar radiation 
field (ISRF) were taken from \cite{Port} which confirmed those from \cite{Chi}.

The same electron spectrum as for the case of IC and the same ISM gas as for the case 
of $\pi^\circ$ gammas have been used to calculate the spectrum of bremsstrahlung gamma 
rays. Figure 1 shows that its contribution does not exceed several percent.

The dominant contribution in the GeV region is from $\pi^\circ$ gamma rays. Inverse 
Compton and bremsstrahlung processes do not contribute more than 5\% of the total gamma
 ray flux. Hence, in order to visualize the possible difference of the shapes for the 
observed and calculated spectra the dominant $\pi^\circ$ spectrum has been increased 
to normalize the total calculated intensity at logE = -1 to the experimental one.    
\subsection{Gamma ray spectral features to be explained} 
It is evident that the calculated (conventional) spectra do not fit the Fermi LAT and 
MILAGRO gamma ray measurements.
The concavity of the measured spectrum at about 10 GeV is stronger than that of the 
calculated spectrum. This feature has been examined by us in \cite{EW6} where it is 
argued that a 'New Component' can be involved at energies below 10 GeV. At this stage 
it can be remarked that the Fermi LAT work \cite{Acke} shows an even more dramatic 
feature just below 10 GeV. Specifically, they appear to show a statistically 
significant change of the fractional intensity difference, (data - model)/data, in 
going from 6 to 9 GeV (~although, as usual, this is presumably diluted by 
systematic errors~). This feature gives support to the idea of a new component but this
 component should cease for energies much above 10 GeV which are mainly the concern of 
the present work. 
  
The Fermi LAT collaboration put forward three possible explanations for the feature 
\cite{Acke}: (i) a contribution of undetected gamma ray point sources: pulsars, pulsar 
wind nebulae, SNRs; (ii) the presence of 'fresh' cosmic ray sources with a harder 
injection spectrum; (iii) different cosmic ray particle spectra in different parts of 
the Galaxy, particularly that in the Inner Galaxy being harder than locally. Having no 
arguments against the possible contribution of undetected gamma ray point sources we 
give stronger support 
to the last two mechanisms (~as mentioned in our earlier work \cite{EW5}~). We argue 
that both of them have the same physical origin - the higher frequency of SN explosions
 in the Inner Galaxy and hence the higher density of SNR.
    
It is known that CR particles measured locally come mostly from a relatively small 
number of CR sources at small distances from the solar system (~mainly within 
$\sim$1kpc~). On the other hand, measured
 gamma rays are produced by CR interacting with the ISM all along the line of
 sight and therefore originating from many more sources. Gamma rays coming from the 
Inner Galaxy contain a larger fraction of those produced in the region of higher
frequency of SN explosions. The higher frequency means that the 'effective' age of SNRs
is younger and therefore the energy spectrum of their produced CR is harder than those 
nearby. The concave shape of the $\pi^0$-gamma spectrum can be due both to the 
contribution of many sources with different slopes of the power law spectra including 
relatively hard ones and the generally harder spectra near to young sources - the 
mechanism proposed in \cite{Qian,EW6} and mentioned as item (ii) in \cite{Acke}. 

In addition to the role of young SNR sources in the Inner Galaxy (~from which CR have 
not diffused very far~) there is the possibility of Inner Galaxy SNR providing flatter 
injection spectra than those elsewhere. Such a result could come from the different ISM
 characteristics, such as magnetic fields, gas density, smaller remnants at
 the time of particle escape, etc. Clearly, if the mean exponent of the injection 
spectrum of CR particles in the Inner Galaxy is reduced a better fit can be achieved. 
A (~not unlikely~) change by $\Delta \gamma = 0.1$, i.e. from $\gamma = 2.15$ to 2.05 
is possible, both by virtue of slightly flatter injection spectra and the increased 
probability of a line of sight penetrating the remnant of a SN before it has had time 
to 'release' its accelerated CR into the ISM.

However, it is seen in Figures 1 that in spite of the fact that our simulated 
spectra show the concavity, followed by the flatter slope at high energies originating 
from this mechanism there is still an excess of the experimental intensities over our 
Monte Carlo simulations, so that we confirm the Fermi LAT conclusion about the 
existence of this excess. The actual reduction of the exponent for particle injection 
spectra needed is greater than $\Delta \gamma = 0.1$ and the explanation of the 
flattening by this mechanism alone requires a more serious modification of the model 
for CR acceleration and propagation, as considered below.
\section{Interpretation in terms of the higher turbulence of the ISM in the Inner 
Galaxy}
\subsection{Cosmic Ray Diffusion}
In our view, a better explanation of the spectral feature is in terms of the effect of
 propagation, as distinct 
from injection, as put forward by us in \cite{EW5}. In that work it was pointed out 
that the exponent of the ambient CR particle spectrum depends on the mode of the 
particle 
diffusion. We are not disputing 'diffusion' as the mechanism responsible for CR 
transport but, rather, its mathematical form. It would be remarkable if this form were 
the same everywhere. In \cite{EW5} it was pointed out that the diffusive properties of 
the ISM depend on the degree of turbulence of the medium which, in turn, depends on the
 position in the Galaxy. This aspect will now be examined in a more detailed way than 
has been done hitherto.
\subsection{Turbulence in the Galaxy} 
Turbulence in the ISM arises from energy input from a variety of sources, such as 
supernovae (SN), SNR, pulsars, stellar winds, jets etc.
A useful summary of the surface densities of 'energy sources' as a function of 
Galactocentric distance is given in \cite{Fate}. In going from 
locally to a Galactocentric distance of 3 kpc the increase in input energy density due 
to the higher density of sources is about 3.

Confirmation that there is an increase in turbulence as one approaches the Galactic 
Centre, as distinct from the energy input, is provided by many indicators. Specific 
ones are the magnetic field \cite{Beck}, where the factor of increase for R = 3 kpc
compared with locally is $\sim$2.8  (~for the total field, regular and irregular~) and 
thermal bremsstrahlung: $\sim$4 \cite{Aber}.
\subsection{Diffusive characteristics as a function of the degree of turbulence}
Ideally, it should be possible to determine the expected value of $\alpha$ as a 
function of the degree of turbulence but there are many factors which make this 
impossible, so far. Nevertheless, the data given in \cite{EW4} allow a very approximate
 result: $\alpha$ changes from 1.0 to 0.5 for a change in turbulent energy by a factor 
3. Thus, we would expect a change in $\alpha$ of about 0.5 in going from locally to 
R = 3 kpc or a change by $\simeq$0.25 averaged over a line of sight through the Inner 
Galaxy (~at $\ell = 0^\circ$~).
\subsection{Determination of the experimental value of $\alpha$} 
Figure 2 gives the result of fitting the Inner Galaxy gamma ray spectrum 
(~$30^\circ < \ell < 65^\circ, |b| < 2^\circ$~) with a reduced value of $\alpha$: 
0.8, such as to give a fit to the data. Although we are probing the Inner Galaxy, 
in fact the closest distance the line of sight gets to the Galactic Centre (~for 
$\ell = 30^\circ$~) is about 4 kpc. The result is that if the value in the true 
Inner Galaxy, at R$\sim$3 kpc were 0.5, that relevant to the present range would be 
$\alpha \simeq 0.9$. Thus, the observed value of $\alpha$ is in the region of 
expectation.
\begin{figure}[ht]
\begin{center}
\includegraphics[height=15cm,width=10cm,angle=-90]{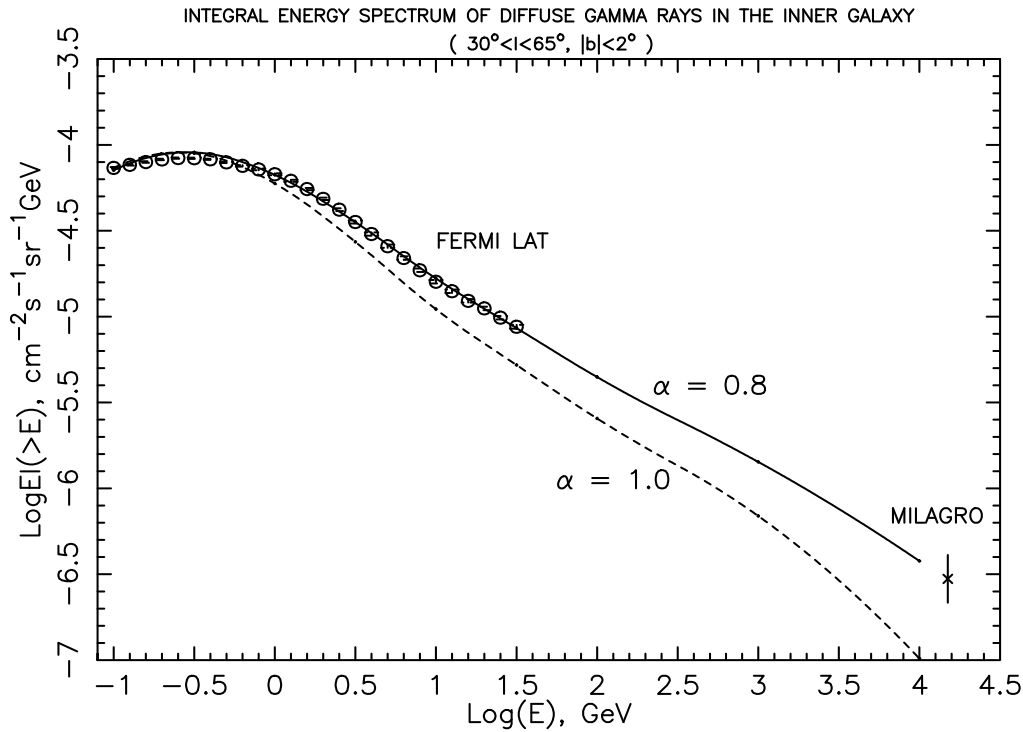}
\end{center}
\caption{\footnotesize Integral energy spectrum of diffuse gamma rays in the Inner 
Galaxy. Comparison of the Fermi LAT (~open circles~) and MILAGRO (~cross~) 
measurements with two versions of
spectra calculated for anomalous diffusion with the parameter $\alpha$ equal to 1.0
(~dashed line~) or 0.8 (~full line~). The agreement of the experimental meaurements 
with the spectrum calculated for the more turbulent ISM in the Inner Galaxy
(~with $\alpha = 0.8$~) is quite good.}
\label{fig:fig3}
\end{figure}

The conversion of the gamma ray energy spectrum observed in the region with Galactic 
coordinates $|\ell| < 80^\circ, |b| < 8^\circ$ to the region of 
$30^\circ < \ell < 65^\circ, |b| < 2^\circ$ described in \S2 and illustrated in Figure 
1 has been made for $\alpha = 1$. The same conversion made using $\alpha = 0.8$ gives 
the spectrum still flatter than that for $\alpha =1$. It makes our conclusion about the
 flatter CR energy spectrum in the Inner Galaxy stronger still. 

At the present stage of the study we do not pretend to specify the precise value of 
$\alpha$. The important point is the qualitative conclusion about the flatter CR energy
 spectrum in the Inner Galaxy than locally, which gives the support to the general 
contention on the non-uniformity of CR characteristics all over the Galaxy. 

 \section{Difference between the gamma-ray spectra in the Inner and Outer Galaxy}
The reduced turbulence in the Outer Galaxy compared with that in the Inner Galaxy leads
 us to predict steeper CR energy spectra in the Outer Galaxy although the wide range of
 latitude (~$|b| < 8^\circ$) means that the far Outer Galaxy is not reached in the used
 Fermi LAT observations. The published Fermi 
LAT measurements cover the area of the whole Outer Galaxy with coordinates 
$80^\circ < \ell < 280^\circ$, $|b| < 8^\circ$ and the collaboration has already 
noticed that the excess of measured intensities above the model calculations at high 
energies is smaller in the Outer compared with the Inner Galaxy. It means that the 
energy spectrum of gamma rays in the Outer Galaxy is indeed steeper than in the Inner
 Galaxy. Figure 3
shows the measured spectra compiled from Figures 15 and 16 of \cite{Acke}. Fitting them
 by power laws at energies above 10 GeV yields values for the differential exponents of
 $2.401 \pm 0.009$ for the Inner Galaxy and $2.525 \pm 0.005$ for the Outer Galaxy. 
Taken at face value the difference is very significant but it is appreciated that 
  systematic errors degrade the results. Presumably some, at least, of the 
systematic errors are the same for the measurements towards, and away from, the 
Galactic Centre so that their effect is not large; nevertheless, it could be 
appreciable. Thus, the conservative statement: 'this supports the prediction of a 
steeper gamma ray spectrum in the Outer Galaxy compared with the Inner' is appropriate.

Both spectra shown in Figure 3 are for the total gamma ray flux including contributions
from discrete sources and the isotropic background. In \S2 it was shown that the 
subtraction of these contributions increases the exponent of the spectrum in the Inner 
Galaxy by (0.117-0.067)/3 = 0.017. It is not enough to remove the 
difference between the calculated exponents of the two spectra in the Inner and Outer 
Galaxy.       
  
\begin{figure}[ht]
\begin{center}
\includegraphics[height=15cm,width=10cm,angle=-90]{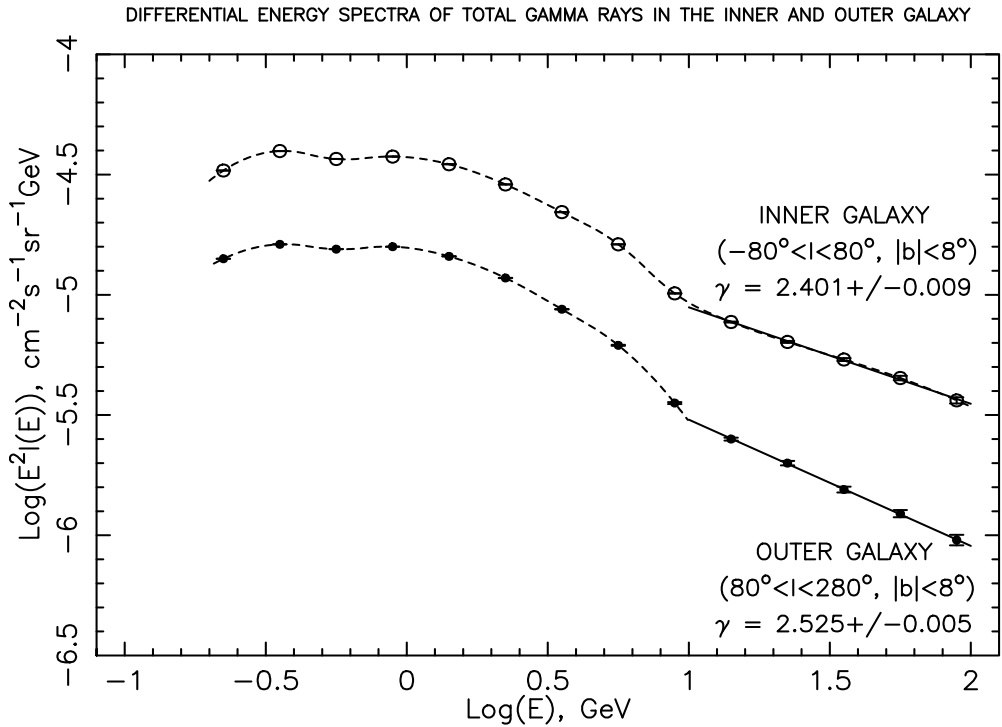}
\end{center}
\caption{\footnotesize Differential energy spectra of total gamma rays in the Inner
Galaxy (~open circles, $|\ell| < 80^\circ, |b| < 8^\circ$~) and in the Outer Galaxy 
(~full circles, $80^\circ < \ell < 280^\circ, |b| < 8^\circ$~) compiled from 
Figures 15 and 16 of \cite{Acke}.The dashed lines are the cubic splines drawn to guide 
the eye. The full lines above $logE = 1$ are best linear fits, the slope of 
which confirms the steeper spectrum in the Outer Galaxy.}    
\label{fig:fig4}
\end{figure}

Inspired by the acceptable agreement of our model calculations with the experimental 
data in the Inner Galaxy shown in Figure 2 we have calculated the gamma ray spectrum 
for the Outer Galaxy  $80^\circ < \ell < 280^\circ$, $|b| < 8^\circ$ at higher energies
 than published by Fermi LAT. The $\alpha$ parameter was taken as 0.8 for the Inner 
Galaxy and 1.0 for the Outer Galaxy. The comparison of these two spectra is shown in 
Figure 4.

 It is seen that the predicted spectrum in the Inner Galaxy at TeV energies is 
considerably flatter than in the Outer Galaxy. Since upper limits for the gamma ray 
intensity in the Outer Galaxy, given by the MILAGRO collaboration are rather high, they
 do not help the comparison.   
\begin{figure}[ht]
\begin{center}
\includegraphics[height=15cm,width=10cm,angle=-90]{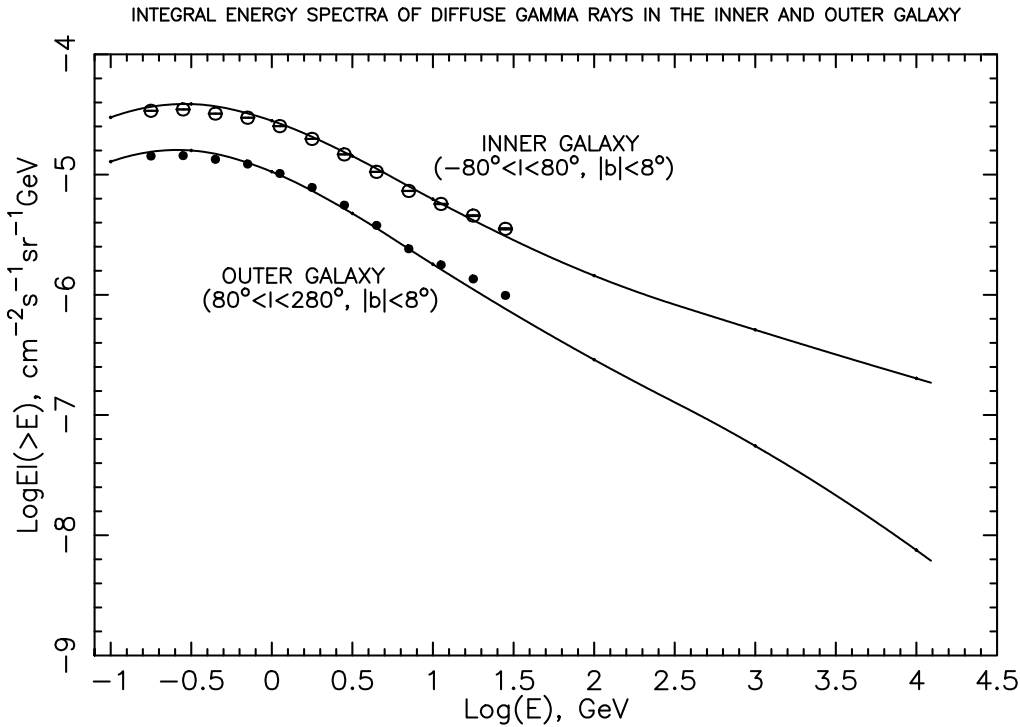}
\end{center}
\caption{\footnotesize Integral energy spectra of diffuse gamma rays in the Inner 
Galaxy (~open circles, $|\ell| < 80^\circ$, $|b| < 8^\circ$~) and in the Outer Galaxy  
(~full circles, $80^\circ < \ell < 280^\circ$, $|b| < 8^\circ$~) from Fermi LAT 
measurements. Upper full curve - calculations for the model of anomalous diffusion with
 $\alpha = 0.8$ in the Inner Galaxy. Lower full curve - calculations for the Outer 
Galaxy with the turbulence in the ISM equal to that locally ($\alpha = 1.0$). It is 
seen that at TeV energies the expected difference between spectra in the Inner and 
Outer Galaxy is higher than at tens of GeV. If the turbulence in the Outer Galaxy is 
smaller than that locally \cite{EW6} then the predicted difference can be bigger still.
 We remind that the 'Outer Galaxy' results refer mainly to the near Outer Galaxy, 
in fact.}   
\label{fig:fig4}
\end{figure}

\section{Comparison with the results of other studies}
The Fermi LAT group themselves \cite{Acke} acknowledge that there is a 
difference in gamma ray spectra between the Inner and Outer Galaxy but their possible 
explanations do not include 'anomalous diffusion'.

In \cite{EW5} we examined the 'low energy' gamma ray spectra from \cite{Hunt}, and used
it to determine the dependence of the proton spectral exponent on Galactocentric
 distance. Although an anomalous excess of gamma rays above several GeV reported by
the EGRET team was not confirmed by the Fermi LAT measurements \cite{Acke} this does 
not invalidate the conclusions. It was found that the differential exponent of the 
gamma ray spectrum increased from $2.40\pm 0.14$ at R = 5kpc to $2.88\pm 0.12$ at 
R = 15kpc. Such a variation is of the order of that for the 'variable turbulence model'
 predictions referred to in \S4.

In that work \cite{EW5} it was shown that there was a simple relationship between the 
anomalous diffusion parameter, $\alpha$, and the ensuing exponent of the differential proton spectrum, $\gamma$. The result was an increase in $\gamma$ of 0.48 for an increase in $\alpha$ (~ie a decrease in the degree of anomaly~) of 0.5.

The MILAGRO group \cite{Abd1,Abd2} interpret the spectral data from their own 
observations for $\ell: 30^\circ - 65^\circ$ in terms of the 'GALPROP' model
 of 'conventional' particle diffusion but with an artificially enhanced IC 
contribution, specifically by increasing the electron intensity by a factor 4. We 
consider that the explanation(s) advanced here are more physical and less ad hoc in the
sense that increased turbulence should have an effect on particle propagation in some 
form, at least. However, a full treatment of the effect has not yet been made..

Finally, reference to be made to a paper \cite{Toma} which appeared after the present 
work had been submitted. The study uses our work \cite{EW5} on anomalous diffusion but 
concentrates on the latitude variation of the CR diffusion properties and examines the 
implications for primary particle spectra, secondary to primary ratios and latitudinal 
anisotropies. Thus, it is complementary to the present paper.

\section{Conclusions and Future Work}
The flattening of the gamma ray spectrum for the Inner Galaxy in comparison with that
 in the Outer Galaxy seems well established (~although, because of the possibility of 
systematic errors, confirmation by independent precision measurements is needed~) and, 
in our view, finds a natural explanation for CR accelerated in SNR in which 
particles are propagated in the increasingly turbulent region encountered as the 
Galactocentric distance is diminished.  

As remarked already (~in \S6~) further work is necessary to quantify the degree of 
anomalous diffusion expected. Further work will include a more detailed spatial 
analysis of the diffusion properties 
of the ISM by way of comparisons of gamma ray spectra from active regions of the Galaxy
 (~such as OB associations~) and at various heights above the Galactic Plane. In terms 
of other galaxies it will be useful to study radio spectra (~from CR electrons and 
magnetic fields~) as a function of galaxy type and luminosities in other wavebands.

Although getting away from the main thrust of the paper, an aspect of potentially 
considerable importance is the role of Gamma Ray 
Astronomy (~via Fermi LAT~) in determining the foreground to be subtracted from the map
 of Cosmic Microwave Background (CMB). In an early work using the 'WMAP' and the EGRET 
gamma ray data (~see \cite{Wibi}~) it was claimed that the 'cleaned' CMB map still 
contained signatures due to CR effects in the Galactic Halo. The relevance to the paper
 is that whereas we have concentrated on large-scale differences in the diffusion 
properties of the Inner and Outer Galaxy there is the likelihood of such differences on
 smaller scales, too. Such changes at higher Galactic latitudes are germane to the CMB 
foreground problem.

\vspace{3mm}

{\bf Acknowledgements} \\

The Kohn Foundation is thanked for supporting this work. We are grateful to Dr. Andrew
Strong for his ready advice and to the referee for numerous helpful suggestions.

\end{document}